# A Monte Carlo Approach to Joe DiMaggio and Streaks in Baseball[1]

Samuel Arbesman
Cornell University
sam.arbesman@cornell.edu

Steven H. Strogatz
Cornell University
shs7@cornell.edu

Abstract

We examine Joe DiMaggio's 56-game hitting streak and look at its likelihood, using a number of simple models. And it turns out that, contrary to many people's expectations, an extreme streak, while unlikely in any given year, is not unlikely to have occurred about once within the history of baseball. Surprisingly, however, such a record should have occurred far earlier in baseball history: back in the late 1800's or early 1900's. But not in 1941, when it actually happened.

---

[1] Some of this work has already appeared in the *New York Times* [5]

The incredible record of Joe DiMaggio in the summer of 1941 is unparalleled. No one has come close—before or since—to equaling his streak of hitting safely in 56 games in a row. People have gone even further and stated that it is the only record in baseball (or perhaps even in all of sports) that never should have happened, statistically speaking: while other records can be explained by expected outliers over the long and varied history of professional baseball (nearly 150 years), DiMaggio's record stands alone [10].

In addition, streaks are of more general interest than the parochial domain of baseball. For example, Bill Miller's Legg Mason Value Trust mutual fund succeeded in beating the S&P 500 for fifteen years in a row [12]. Streaks clearly have a certain correspondence in relation to skill, and teasing out the difference between luck and skill is important. Therefore, a better understanding of streaks is a useful exercise.

We wondered whether Joe DiMaggio's record was more likely than people might think. To study this, we constructed a series of simple Monte Carlo simulations, using a comprehensive baseball statistics database from 1871 to 2004 [11]. The first and simplest model is as follows:

Each player, for each season that he played, was characterized by a few numbers: the number of games played, number of plate appearances, and number of hits in the season. Number of plate appearances is the sum of at-bats, times walked, being hit by a pitch, sacrifice hits, and any other way in which a player appears at the plate and either has a chance for a hit, or is denied that chance (such as being walked or being hit by a pitch). Our data contained nearly all of the categories for plate appearances, so the values used are as similar as possible to the actual number of plate appearances in an entire season. Then, we calculated each player's probability of getting at least one hit in any game in which he played for that year as follows:

$M$ = number of games
$n$ = number of plate appearances per game = number of plate appearances/$M$
$p$ = number of hits/number of plate appearances

The probability of not getting a hit for a single plate appearance is 1-$p$. Assuming that successive plate appearances within a single game are independent, the probability of not getting a hit for an entire game is then $(1-p)^n$. So, the probability of getting at least one hit in a single game is $1-(1-p)^n = g$. Similar calculations were done in a paper by Michael Freiman [8].

Given $g$ for all players and for all seasons (this was only calculated for players who had at least 300 or 450 plate appearances in the season, as seen below), we can simulate a player's streaks for a season. We randomly generate $G$ games, and count how many successful games in a row there were. This is done for all players and for all seasons, effectively simulating an entire baseball history. We tabulate the long streaks in each history, including the longest streak, when it occurred, and who had it. Of course, the actual probabilities for streaks of various lengths can be calculated exactly for this simple model, but we later modify the model in various ways, so a computational approach is most reasonable [13]. To gain reliable statistics, we performed these simulations 10,000 times.

For this model, a streak of 56 games or longer occurs about 49% of the time. Figure 1 shows the histogram of the longest streaks, in each of the 10,000 simulated histories of baseball. In this and all other instances in the paper, we take an occurrence of more than 5% in our simulations—the conventional number in hypothesis testing—to be indicative that this could happen by chance reasonably often.

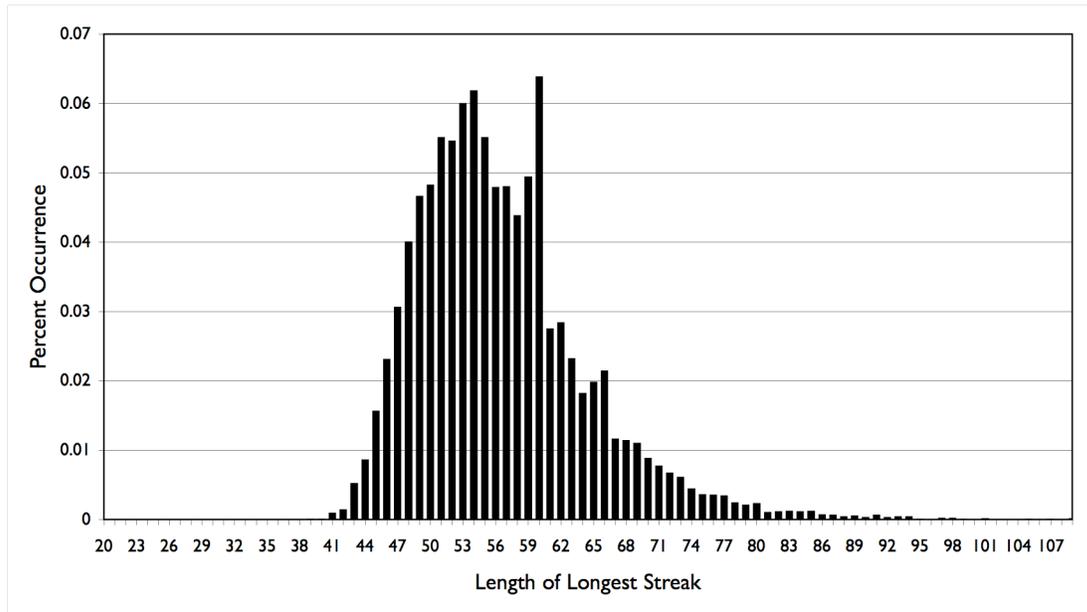

Figure 1. Distribution of the longest streaks in 10,000 simulated runs of baseball history. The curve is roughly bell-shaped with a longer tail to the right, favoring longer streaks. The most frequent outcome was a record streak of 60 games, but this is due to a short, high impact season of Ross Barnes in 1873. Adding up all the heights of the columns for streaks 56 games or longer yields about 49% of all the data, meaning that streaks as long as DiMaggio's are expected about 49% of the time.

The most frequent outcome was a record streak of 60 games. This is due to a single player in a single season: Ross Barnes in 1873. Ross Barnes, who had an incredibly high probability of getting at least one hit in any game (0.946) only played 60 games that season. These two facts combined to yield a large number of simulated histories where Barnes hit in every game in the entire season, giving us the spike at 60.

The above model, which we shall term Model A (one with constant probabilities), overestimates certain long streaks, as exemplified by Ross Barnes acting as an outlier. So, Model A was then modified to improve its robustness and its similarity to actual baseball history. We added game-to-game variation to $g$, due to variation in number of plate appearances in games and opposing pitching ability. The true amount of variation due to pitching ability or variable number of plate appearances is more complicated to estimate, especially due to a lack of more detailed game-by-game data for the entire dataset, as discussed later. So, for a rough approach, we ran the simulations with a 10% and 20% uniform variation in *g* for each game. The 10% variation model is Model B, and the 20% variation model is Model C.

Table 1. Model names and descriptions.

| Model | Description |
|---|---|
| Model A | Constant *g* and a minimum of 300 plate appearances in a season for inclusion |
| Model B | *g* varied by up to 10% and a minimum of 300 plate appearances in a season for inclusion |
| Model C | *g* varied by up to 20% and a minimum of 300 plate appearances in a season for inclusion |
| Model B (450 PA cutoff) | *g* varied by up to 10% and a minimum of 450 plate appearances in a season for inclusion |
| Model B (post-1905) | *g* varied by up to 10% and a minimum of 300 plate appearances in a season for inclusion, and only data from 1905 onwards was used |

To test the validity of all of these models, we compared their predictions against the observed number of shorter streaks in baseball history. The distribution of streaks of length 30 or greater follows what is approximately an exponential distribution [1], as shown in Figures 2 and 3.

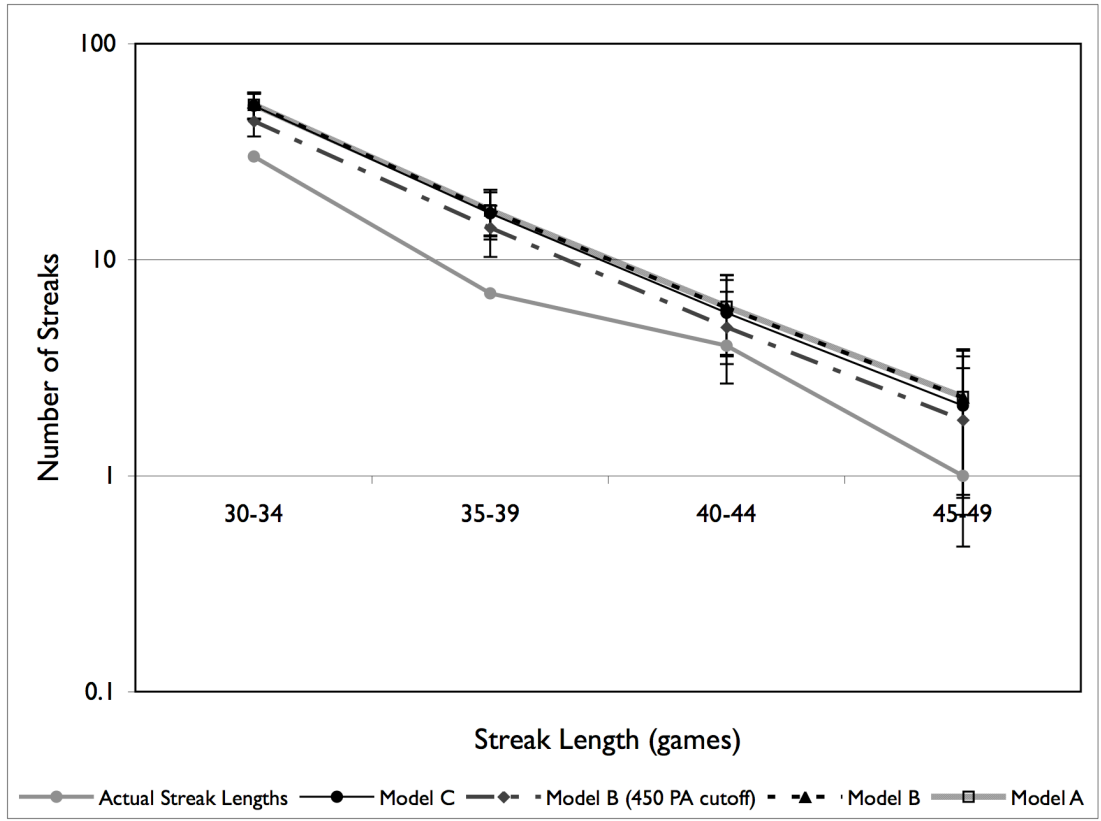

Figure 2. The distribution of streaks (aside from DiMaggio's) for all of baseball to the present [1]. The following models are included: Model B, using all available data and a minimum number of 300 plate appearances in a season; Model B, using all available data and a minimum number of 450 plate appearances in a season; and Model C, using all available data and a minimum number of 300 plate appearances in a season. Models A, B, and C all yield extremely similar results and overlap quite a bit in the figure.

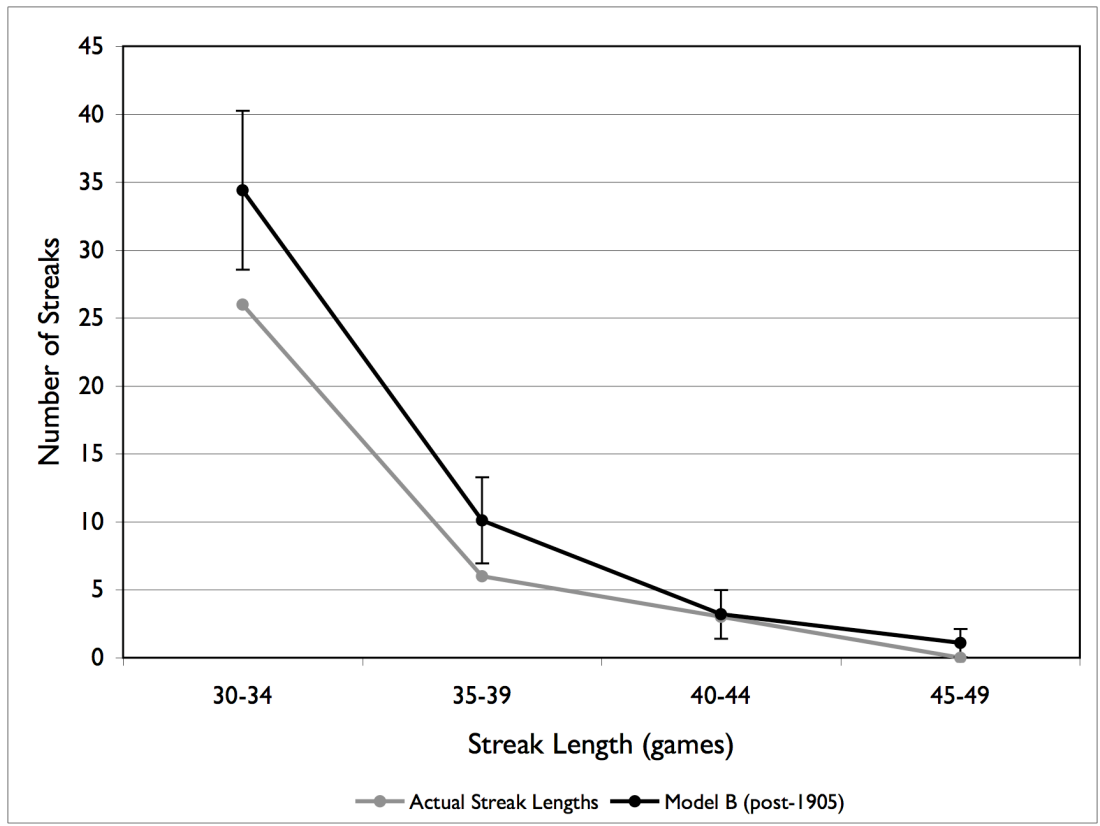

Figure 3. The distribution of streaks (aside from DiMaggio's) from 1905 to the present [1]. Model B, using data only from 1905 until 2005, is included for comparison.

It is found that the simulated histories also have exponential distributions for the length of the streaks. The slope varies as variation is added, the minimum number of plate appearances to be included in the simulation is increased, or the number of years being examined is limited. However, it can be seen that the results are similar to what is found in 'real' baseball, and that the functional form is preserved by these simple models.

Specifically, all the Models are within a single standard deviation for the number of actual streaks of lengths 40-44 and 45-49. For the shorter streaks, the models overestimate, by about 50%, yielding same order-of-magnitude results. These might be improved by refining the model, as discussed later.

Furthermore, adding variation turns out to change very little of the results, and actually increases the correspondence between the streak length data and the simulations, without changing the probability of outlier streaks, such as DiMaggio's, appreciably.

The true amount of variation due to pitching ability or variable number of plate appearances is not known, however it could be estimated. One method to examine both of these is to use box score data, which is available in an electronic format for the past fifty years or so [2]. By examining the variation in at-bats, we could get a better estimate of plate appearance variation. In addition, by examining the variation in hitting ability against different pitchers, we could estimate the range for hitting ability for an average player. In this case though, the simulations were run with a 10% and 20% uniform variation in *g* for each game, as an approximation. As seen in Table 2, the probability of a DiMaggio streak was still non-trivial.

In addition, since, as can be seen from Figure 1 and 4, there are certain spikes due to outliers in the early days of the game, a simulation using data only for 1905 and later was conducted. This is considered to be the more modern era of baseball. Using this more limited data sample, the spike of 60 game streaks due to Ross Barnes in the early 1870's is eliminated and the curve becomes smoother (Figure 5). Streaks become less likely, but DiMaggio-like streaks (56 games or longer) still occur nearly 20% of the time. Please see Table 2 and Figures 4-7 for all results.

Table 2. Probabilities of DiMaggio-like Streaks in a variety of models. All demonstrate a non-trivial likelihood of such an extreme streak.

|  | *Model A* | *Model B* | *Model B (post-1905)* | *Model C* | *Model B (450 PA min.)* |
|---|---|---|---|---|---|
| *Probability of DiMaggio-Like Streak* | 0.49 | 0.49 | 0.18 | 0.39 | 0.38 |

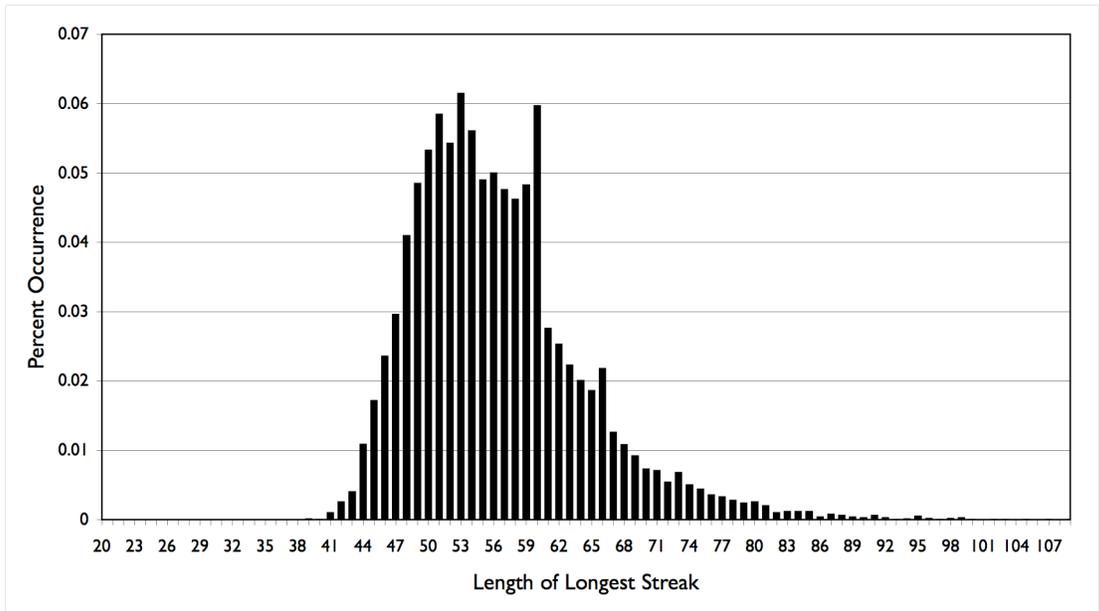

Figure 4. Streak length histogram for Model B.

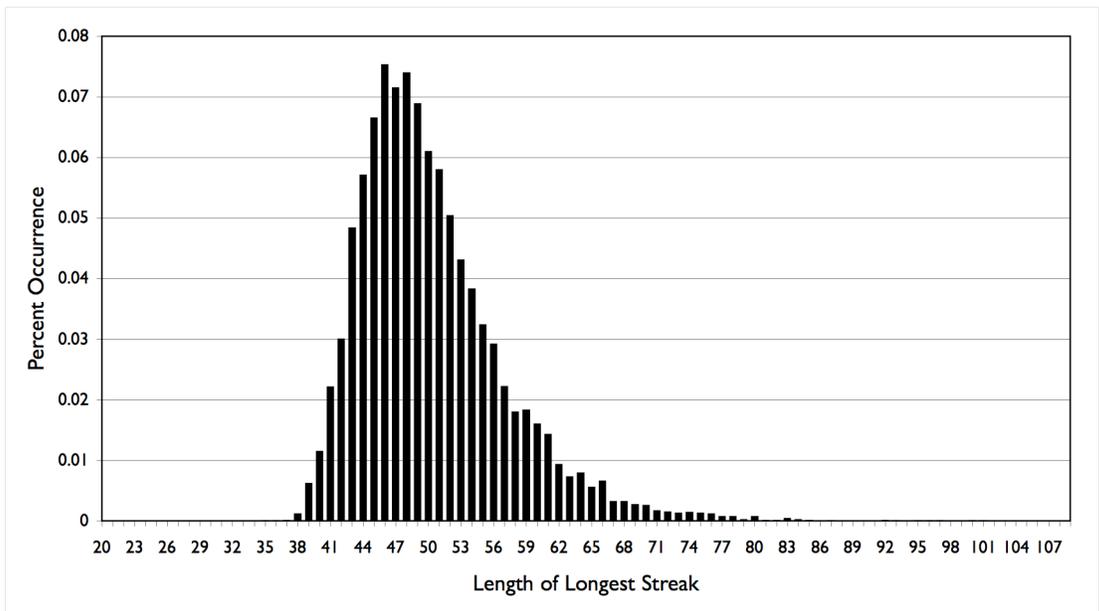

Figure 5. Streak length histogram for Model B (1905 or later).

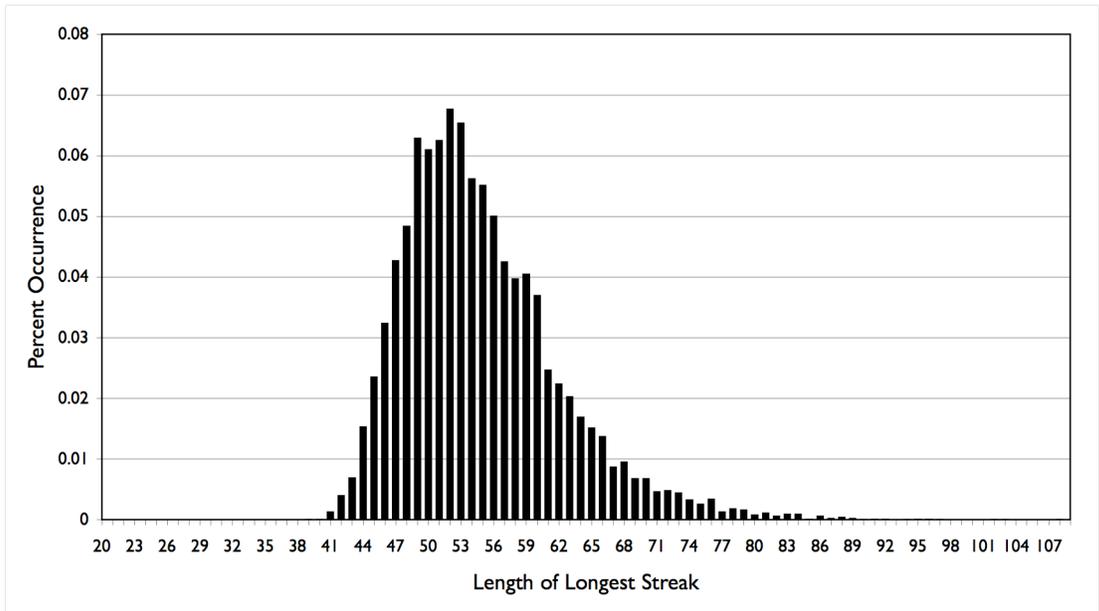

Figure 6. Streak length histogram for Model C.

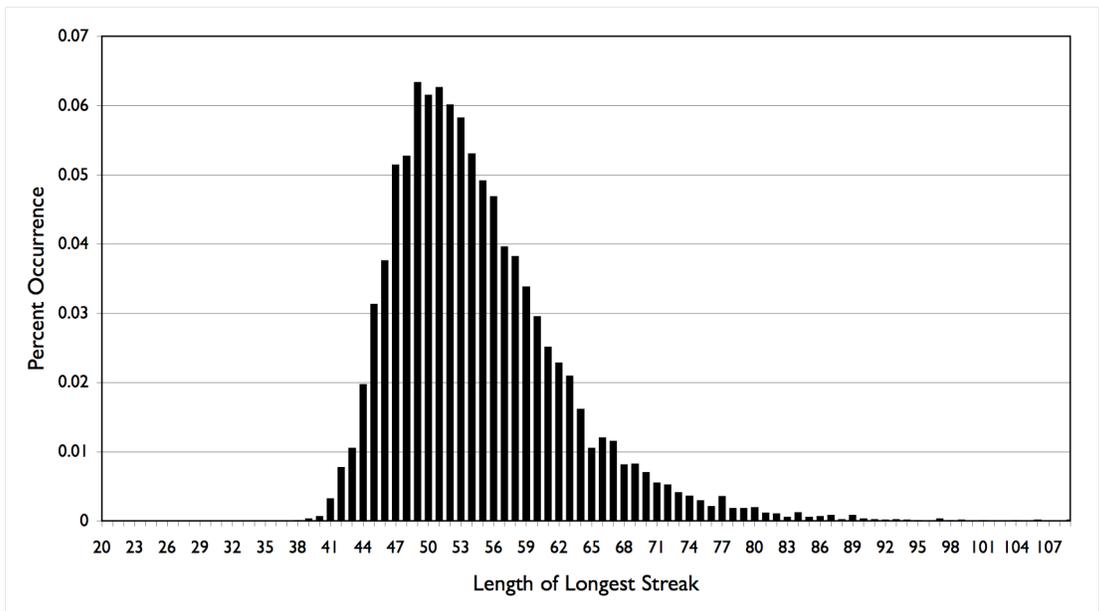

Figure 7. Streak length histogram for Model B (450 plate appearance minimum).

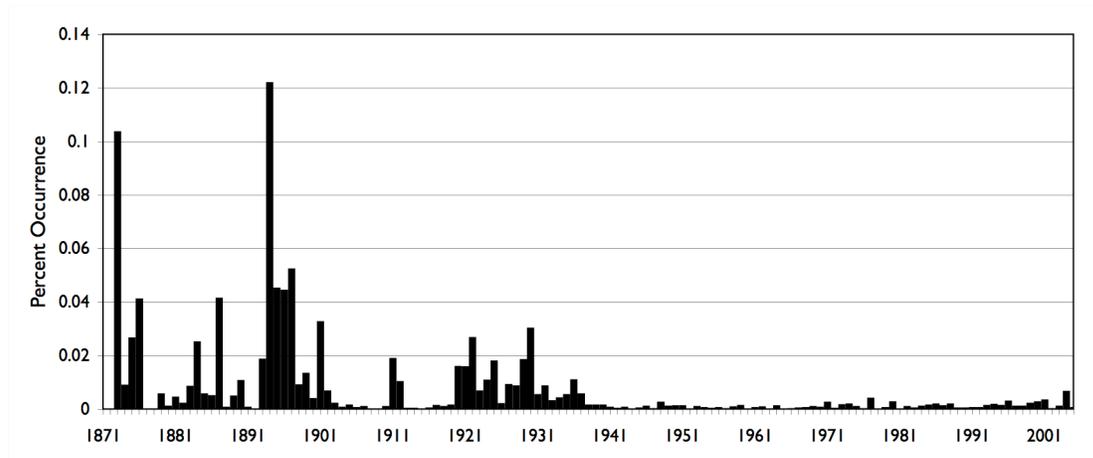

Figure 8. Timeline for streak likelihood (Model A). This graph shows when the longest streaks occurred in our simulations. The record was likeliest to have been set in 1894, when Hugh Duffy batted .440 and had a 91 percent chance of hitting successfully in each game. Other likely periods for the record to be set were in the early teens (Ty Cobb's era) and throughout the 20's and 30's. Joe DiMaggio's miracle year of 1941 was a relative dry spell, making his achievement all the more stunning. The other models have relatively similar yearly histograms.

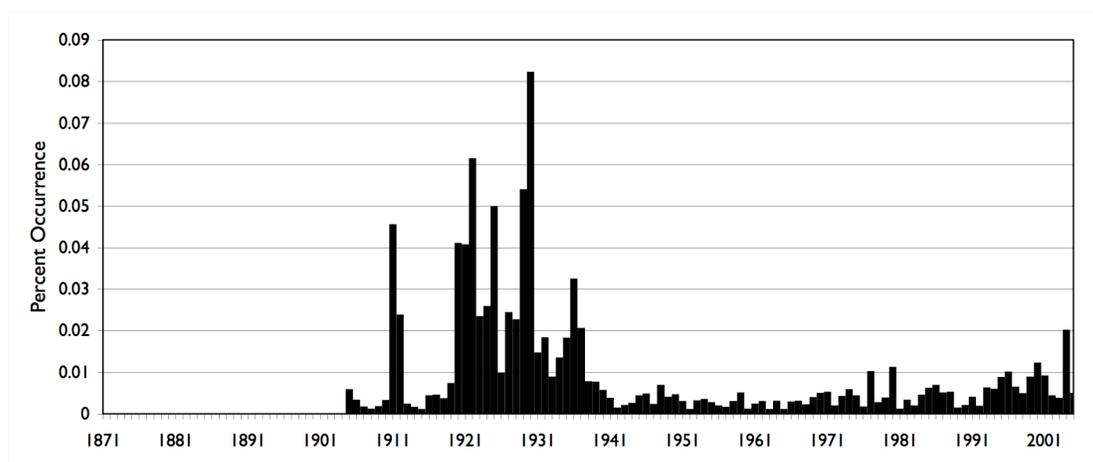

Figure 9. Timeline for streak likelihood (Model B, post-1905).

The surprising thing found in the simulation is when DiMaggio's streak occurred, as seen in Figures 8 and 9. It should have occurred far earlier in the history of baseball, back in the late 1800's or early 1900's. But not in 1941, or similar years. Years with a similar probability of 1941 or worse of holding the record account for only about 5% of the records (Model A).

Table 3. The ten players who are likeliest to hold the record for longest streak, for each model. The fraction of the simulations they each hold the streak is included, as *P(Streak)*.

**Model A**

| Player | P(Streak) |
|---|---|
| Ross Barnes | 0.1035 |
| Willie Keeler | 0.0646 |
| Hugh Duffy | 0.048 |
| Jesse Burkett | 0.0465 |
| Sam Thompson | 0.0368 |
| Ed Delahanty | 0.0335 |
| Nap Lajoie | 0.0333 |
| George Sisler | 0.0291 |
| Ty Cobb | 0.0286 |
| George Wright | 0.0284 |

**Model B (post-1905)**

| Player | P(Streak) |
|---|---|
| George Sisler | 0.0723 |
| Ty Cobb | 0.0685 |
| Rogers Hornsby | 0.0395 |
| Al Simmons | 0.0386 |
| Bill Terry | 0.0288 |
| Paul Waner | 0.0217 |
| Ichiro Suzuki | 0.0217 |
| Chuck Klein | 0.0217 |
| Joe Jackson | 0.021 |
| Harry Heilmann | 0.0197 |

**Model B**

| Player | P(Streak) |
|---|---|
| Ross Barnes | 0.0905 |
| Willie Keeler | 0.0651 |
| Hugh Duffy | 0.0488 |
| Jesse Burkett | 0.0448 |
| Sam Thompson | 0.0411 |
| Ed Delahanty | 0.0363 |
| George Sisler | 0.0311 |
| Nap Lajoie | 0.0305 |
| Ty Cobb | 0.0295 |
| George Wright | 0.0288 |

**Model B (450 PA Cutoff)**

| Player | P(Streak) |
|---|---|
| Willie Keeler | 0.0806 |
| Jesse Burkett | 0.0567 |
| Hugh Duffy | 0.0566 |
| Ed Delahanty | 0.0494 |
| Nap Lajoie | 0.0389 |
| Tip O'Neill | 0.0388 |
| George Sisler | 0.0382 |
| Ty Cobb | 0.0357 |
| Sam Thompson | 0.0338 |
| Al Simmons | 0.0219 |

**Model C**

| Player | P(Streak) |
|---|---|
| Willie Keeler | 0.0512 |
| Ross Barnes | 0.0458 |
| Jesse Burkett | 0.0431 |
| Sam Thompson | 0.037 |
| Hugh Duffy | 0.0343 |
| George Sisler | 0.0339 |
| Ed Delahanty | 0.0338 |
| Nap Lajoie | 0.0308 |
| Ty Cobb | 0.0296 |
| Al Simmons | 0.0207 |

On the other hand, while DiMaggio is not the likeliest person to hold the record (he is over forty-seventh most likely in Model A), there is not a single likely player. As seen in Table 3, the top three players—Ross Barnes, Willie Keeler, and Hugh Duffy—only account for about 21.6% of the record-holding streaks in our simulations, even in Model A, which overestimates the likelihood (especially for Ross Barnes). And players of DiMaggio's caliber or less account for nearly 25% of the streaks.

So, while no single player is especially likely to hold the record, it is likely that an extreme streak would have occurred. This subtle probabilistic concept has been discussed by Diaconis and Mosteller within the context of calculating the probability of a double-lottery winner [6]. While the probability of a certain individual winning the lottery twice is extremely small, the fact that someone somewhere will do this (since there are many people who buy lottery tickets on a regular basis) is virtually assured. Diaconis and Mosteller like to think about this in terms of their law of truly large numbers: "With a large enough sample, any outrageous thing is likely to happen."

Richard Feynman once made a similar point. He walked into a lecture hall and noted that he just saw the most amazing thing: a car with the license plate ANZ 912; what are the odds of that? [7]. In our model, while there are some players more likely to hold the streak record, there is a 25% chance that the 'unlikely' players (DiMaggio's likelihood or below) are the ones with the record.

Lastly, we checked the distribution of the difference in length between the longest streak in a simulation, and the second-longest streak. Stephen Jay Gould, in his article, 'The Streak of Streaks', notes that 'DiMaggio's fifty-six–game hitting streak is ridiculously and almost unreachably far from all challengers (Wee Willie Keeler and Peter Rose, both with forty-four, come second)' [10]. Is it true that DiMaggio's streak is much farther away from the other streaks than we might expect, or is this a red herring?

Using Models A and B (post-1905) as representative of the models, we checked to see if Gould's thinking was correct. And it turns out that in Model A, nearly 20% of the simulations had a difference between the longest and second-longest streak of 12 games or more. A histogram of the differences is shown in Figure 10. For Model B (post-1905), the number was 13% of the simulations for a difference of 12 games or more. So, DiMaggio as an apparent outlier is not actually true, and this model is able to replicate the observed large differences in streak length.

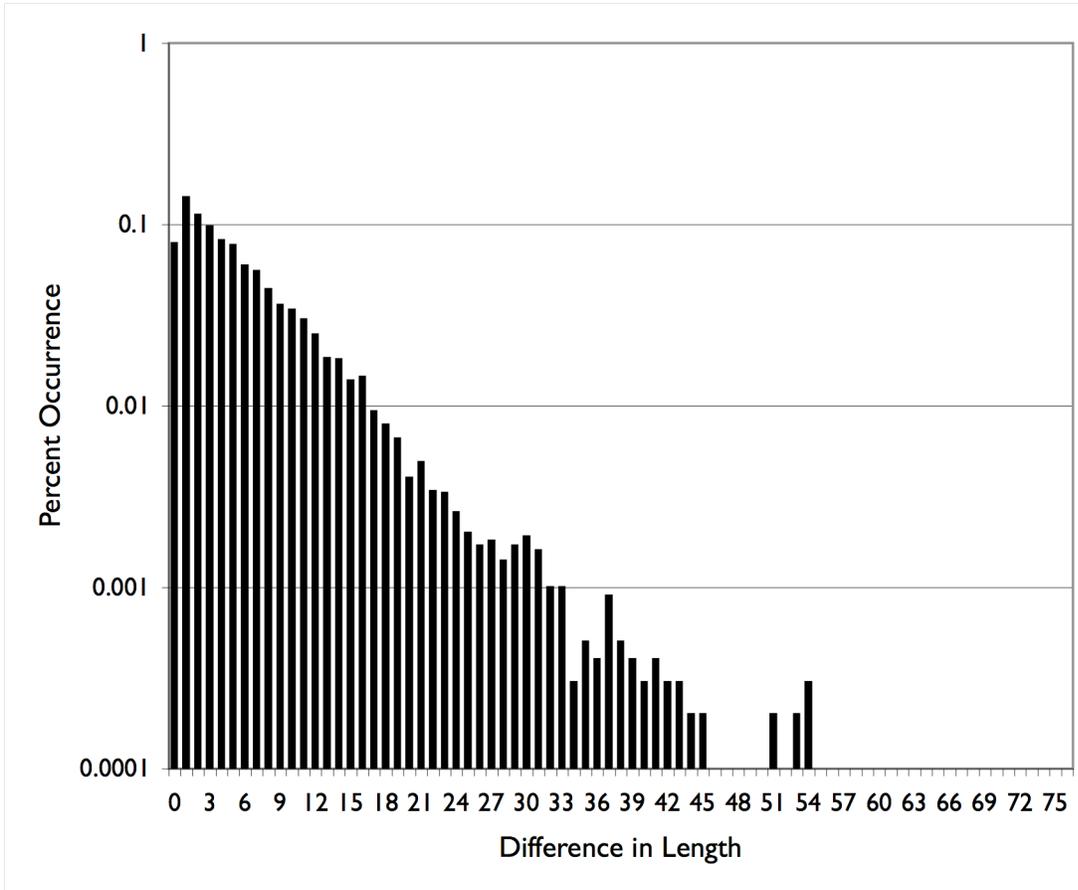

Figure 10. Histogram in difference in length between the longest streak and the second-longest streak for Model A. The y-axis is logarithmic to make it clear that the difference in length appears to follow an exponential distribution (hence a linear decay on a log-linear graph). The mean difference in length is 6.16.

While an effective model and one that provides surprising results, this is of course still an extremely crude model for baseball. Adding the factor of opposing team abilities would presumably change the results, though it is unclear how. There is a model that has recently been developed which looks at streaks of wins between opposing teams in baseball, and this would be a reasonable model to adapt for pairs of pitchers and batters [15].

Furthermore, examination of the psychological effects of a streak has to be taken into account. For example, Roger Maris's hair fell out in large chunks while chasing Babe Ruth's home run record [4]. Stress can certainly affect you.

This is related to the issue of the independence assumption, that is, whether or not a hit in a game is independent of the games before it. Trent McCotter has attempted to examine this, by taking the time series of games and asking whether or not there was a hit in the game [14]. This yields a series of successes and failures. To test the independence assumption, he shuffled the order of the successes to see if the randomized streaks are longer or shorter than expected. He found that the randomized streaks are somewhat shorter than expected, arguing against independence in the real data and hinting at a psychological component or some other factor. Intriguingly, this is different than what has been found in relation to the Hot Hand phenomenon in basketball [9], or even an analysis of successful at-bats [3], which in both cases yielded streaks indistinguishable from random chance.

However, if this model is a moderately realistic one, and it seems to be, then it provides a statistically informed check to our collective baseball intuition. Joltin' Joe's record, while certainly incredible, is in fact not that unlikely within the long history of baseball. But that he did it is certainly still achievement indeed.

Acknowledgements

We thank Michael Mauboussin for raising the issue that initially stimulated our research, and Andy Ruina, Trent McCotter, Stuart Rojstaczer, Tim Novikoff, and so many others for their helpful comments on our op-ed piece [5]. Their insights have helped us improve the models and analysis presented here.